\documentclass[sigconf,review=False]{acmart}

\usepackage{array}
\usepackage{graphicx}
\usepackage{clrscode}
\usepackage{balance}
\usepackage{subfigure}
\usepackage{multirow}
\usepackage{booktabs}
\usepackage{adjustbox}
\usepackage[justification=justified,skip=5pt]{caption}
\usepackage{float}
\usepackage{color}
\usepackage{soul}
\usepackage{mdframed}
\usepackage{amsopn}
\usepackage{mathrsfs}
\usepackage{mathtools}
\usepackage{amsmath}
\usepackage{arydshln}
\usepackage{hyperref}
\usepackage{multicol}
\usepackage{blkarray}
\usepackage{enumerate}
\usepackage{courier}
\usepackage{rotating}
\usepackage{booktabs}
\usepackage{diagbox}
\usepackage{fancybox}
\usepackage{minibox}
\usepackage{cases}
\usepackage{bm}
\usepackage[lined, ruled, commentsnumbered, linesnumbered]{algorithm2e} %
\usepackage{enumitem}
\setlist[itemize]{leftmargin=*}

\DeclareMathOperator*{\argmax}{arg\,max}

\AtBeginDocument{%
  \providecommand\BibTeX{{%
    \normalfont B\kern-0.5em{\scshape i\kern-0.25em b}\kern-0.8em\TeX}}}

\copyrightyear{2021} 
\acmYear{2021} 
\setcopyright{acmcopyright}
\acmConference[WSDM '21]{Proceedings of the Fourteenth ACM International Conference on Web Search and Data Mining}{March 8--12, 2021}{Virtual Event, Israel}
\acmBooktitle{Proceedings of the Fourteenth ACM International Conference on Web Search and Data Mining (WSDM '21), March 8--12, 2021, Virtual Event, Israel}
\acmPrice{15.00}
\acmDOI{10.1145/3437963.3441824}
\acmISBN{978-1-4503-8297-7/21/03}

\begin{document}
\fancyhead{}
% \title[Fairness Constrained Policy Optimization for Recommender Systems]{Fairness Constrained Policy Optimization for\\ Recommender Systems}
\title[Towards Long-term Fairness in Recommendation]{Towards Long-term Fairness in Recommendation}

\author[Ge et al.]{Yingqiang Ge$^{\dagger}$, Shuchang Liu${^\dagger}$, Ruoyuan Gao$^{\dagger}$, Yikun Xian${^\dagger}$, Yunqi Li$^{\dagger}$, Xiangyu Zhao${^\ast}$,}
\author{Changhua Pei${^\ddagger}$, Fei Sun${^\ddagger}$, Junfeng Ge${^\ddagger}$, Wenwu Ou${^\ddagger}$, Yongfeng Zhang${^\dagger}$}
\affiliation{%
  \institution{$^{\dagger}$Rutgers University \qquad ${^\ast}$ 
Michigan State University \qquad $^{\ddagger}$Alibaba Group}
}
\email{{yingqiang.ge, sl1471, ruoyuan.gao, yikun.xian, yunqi.li}@rutgers.edu, zhaoxi35@msu.edu, 
changhuapei@gmail.com,}
\email{ofey.sunfei@gmail.com, beili.gjf@alibaba-inc.com,
santong.oww@taobao.com, 
yongfeng.zhang@rutgers.edu}

\renewcommand{\authors}{Yingqiang Ge, Shuchang Liu, Ruoyuan Gao, Yikun Xian, Yunqi Li, Xiangyu Zhao, Changhua Pei, Fei Sun, Junfeng Ge, Wenwu Ou, Yongfeng Zhang}

\begin{abstract}
As Recommender Systems (RS) influence more and more people in their daily life, the issue of fairness in recommendation is becoming more and more important.
Most of the prior approaches to fairness-aware recommendation have been situated in a static or one-shot setting, where the protected groups of items are fixed, and the model provides a one-time fairness solution based on fairness-constrained optimization. 
This fails to consider the dynamic nature of the recommender systems, where attributes such as item popularity may change over time due to the recommendation policy and user engagement. 
For example, products that were once popular may become no longer popular, and vice versa. 
As a result, the system that aims to maintain long-term fairness on the item exposure in different popularity groups must accommodate this change in a timely fashion.

Novel to this work, we explore the problem of long-term fairness in recommendation and accomplish the problem through dynamic fairness learning. 
We focus on the fairness of exposure of items in different groups, while the division of the groups is based on item popularity, which dynamically changes over time in the recommendation process.
We tackle this problem by proposing a fairness-constrained reinforcement learning algorithm for recommendation, which models the recommendation problem as a Constrained Markov Decision Process (CMDP), so that the model can dynamically adjust its recommendation policy to make sure the fairness requirement is always satisfied when the environment changes.
Experiments on several real-world datasets verify our framework's superiority in terms of recommendation performance, short-term fairness, and long-term fairness.

% \rg{(we propose the long-term fairness problem as a new problem, or we propose to address this problem with a novel approach?)}
% Consequential recommendation decisions will reshape the population over time, as the current recommendation policy will change the exposure distribution of underlying items, which will result in the dynamic changes of fairness (as group members will change).
% trade-off between recommendation performance and fairness performance not only from a short-term perspective, but also a long-term perspective.
% \ykcomment{shall we mention ``dynamic'' in the title?}
% \rg{dynamic to achieve long-term?}
\end{abstract}





\maketitle

\section{Introduction}
% The rapidly evolving e-commerce touches almost every aspect of our lives. 
% We now turn to Amazon for a daily necessity, LinkedIn for a job, and Uber for a ride. 
% And the 
Personalized recommender system (RS) is a core function of many online services such as e-commerce, advertising, and online job markets.
Recently, several works have highlighted that RS may be subject to algorithmic bias along different dimensions, leading to a negative impact on the underrepresented or disadvantaged groups \cite{Geyik2019,Zhu2018,fu2020fairness,singh2018fairness,ge2020understanding}.
For example, the ``Matthew Effect'' becomes increasingly evident in RS, where some items get more and more popular, while the long-tail items are difficult to achieve relatively fair exposure \cite{10.1145/3292500.3330707}.
Existing research on improving fairness in recommendation systems or ranking has mostly focused on static settings, which only assess the immediate impact of fairness learning instead of the long-term consequences \cite{doi:10.1177/1064804619884160, ijcai2019-862}.
For instance, suppose there are four items in the system, A, B, C, and D, with A, B belonging to the popular group $G_0$ and C, D belonging to the long-tail group $G_1$.
When using demographic parity as fairness constraint in recommendation and recommend two items each time, without considering the position bias, we will have AC, BC, AD, or BD to be recommended to consumers.
Suppose D has a higher chance of click,
% is an item with lots of potential, as long as a consumer sees D, he/she is willing to click on it.
then after several times, D will get a higher utility score than other items, but since D is still in $G_1$, the algorithm will tend to recommend D more to maximize the total utility and to satisfy group fairness. 
This will bring a new ``Matthew Effect'' on $G_1$ in the long term.
The above example shows that imposing seemingly fair decisions through static criteria can lead to unexpected unfairness in the long run.
In essence, fairness cannot be defined in static or one-shot setting without considering the long-term impact, and long-term fairness cannot be achieved without understanding the underlying dynamics.

We define \textit{static fairness} as the one that does not consider the changes in the recommendation environment, such as the changes in item utility, attributes, or group labels due to the user feedback/interactions throughout the recommendation process. 
Usually, static fairness provides a one-time fairness solution based on fairness-constrained optimization. 
\textit{Dynamic fairness}, on the other hand, considers the dynamic factors in the environment and learns a strategy that accommodates such dynamics.
Furthermore, \textit{long-term fairness} views the recommendation as a long term process instead of a one-shot objective and aims to maintain fairness in the long run by achieving dynamic fairness over time.

Technically, we study the long-term fairness of item exposure in recommender systems, while items are separated into groups based on item popularity. 
% (total number of exposure for an item), i.e. popular item group $G_0$ and long-tail item group $G_1$.
The challenge is that during the recommendation process, items will receive different extents of exposure based on the recommendation strategy and user feedback, causing the underlying group labels to change over time.
% and the members in each group will begin to transfer since some items may no longer remain popular, and some long-tail items may become more and more popular.
% We define these transitional processes as the dynamics of fairness of item exposure.
% The simplest way to address this is to solve the whole constrained optimization problem after each recommendation. 
% However, this process is significantly time-consuming since most approaches treat fairness constrained optimization as a linear program.
To achieve the aforementioned long-term fairness in recommendation, we pursue to answer the following three key questions:
\begin{itemize}
    \item How to model long-term fairness of item exposure with changing group labels in recommendation scenarios?
    \item How to update the recommendation strategy according to real-time item exposure records and user interactions?
    \item How to optimize the strategy effectively over large-scale datasets?
\end{itemize}
% In order to address the aforementioned dynamics in recommendation fairness, in this work, we leverage constrained reinforcement learning to 

% In this work, we study the problem of long-term fairness in recommendation aiming to address the above challenges simultaneously.
In this work, we aim to address the above challenges simultaneously.
Specially, we propose to model the sequential interactions between consumers and recommender systems as a Markov Decision Process (MDP), and then turn it into a Constrained Markov Decision Process (CMDP) by constraining the fairness of item exposure at each iteration dynamically.
We leverage the Constrained Policy Optimization (CPO) with adapted neural network architecture to automatically learn the optimal policy under different fairness constraints.
We illustrate the long-term impact of fairness in recommendation systems by providing empirical results on several real-world datasets, which verify the superiority of our framework on recommendation performance, short-term fairness, and long-term fairness.
To the best of our knowledge, this is the first attempt to model the dynamic nature of fairness with respect to changing group labels, and to show its effectiveness in the long term.
% in terms of the trade-off between recommendation performance and fairness performance not only from a short-term perspective, but also a long-term perspective.

% ACM's consolidated article template, introduced in 2017, provides a
% consistent \LaTeX\ style for use across ACM publications, and
% incorporates accessibility and metadata-extraction functionality
% necessary for future Digital Library endeavors. Numerous ACM and
% SIG-specific \LaTeX\ templates have been examined, and their unique
% features incorporated into this single new template.

% If you are new to publishing with ACM, this document is a valuable
% guide to the process of preparing your work for publication. If you
% have published with ACM before, this document provides insight and
% instruction into more recent changes to the article template.

% The ``\verb|acmart|'' document class can be used to prepare articles
% for any ACM publication --- conference or journal, and for any stage
% of publication, from review to final ``camera-ready'' copy, to the
% author's own version, with {\itshape very} few changes to the source.

\section{Related Work}
\subsection{Fairness in Ranking and Recommendation}
There have been growing concerns on fairness recently, especially in the context of intelligent decision-making systems, such as recommender systems. Various types of bias have been found to exist in recommendations such as gender and race~\cite{chen2018investigating,abdollahpouri2019unfairness}, item popularity~\cite{Zhu2018}, user feedback~\cite{fu2020fairness} and opinion polarity~\cite{Yao2017}. Different notions of fairness and algorithms have since been proposed to counteract such issues. There are mainly two types of fairness definitions in recommendations: \textit{individual} fairness and \textit{group} fairness. The former requires treating individuals similarly regardless of their protected attributes, such as demographic information, while the latter requires treating different groups similarly. Our work focuses on the popularity group fairness, yet also addresses individual fairness through accommodation to dynamic group labels.  

The relevant methods related to fairness in ranking and recommendation can be roughly divided into three subcategories: optimizing utility (often represented by relevance) subject to a bounded fairness constraint \cite{
% celis2018algorithmic,
% biega2018equity, 
singh2018fairness,Geyik2019,zehlike2017fa,celis2018ranking}, 
optimizing fairness with a lower bound utility \cite{Zhu2018}, and jointly optimizing utility and fairness \cite{celis2019controlling}. 
Based on the characteristics of the recommender system itself, there also have been a few works related to multi-sided fairness in multi-stakeholder systems ~\cite{burke18a,mehrotra2018,Gao2019how}. These works have proposed effective algorithms for fairness-aware ranking and recommendation, yet they fall in the category of \textit{static fairness} where the protected attribute or group labels were fixed throughout the entire ranking or recommendation process. Therefore, it is not obvious how such algorithms can be adapted to dynamic group labels that change the fairness constraints over time. The closest literature to our work on \textit{dynamic fairness} includes \citeauthor{Saito20} ~\cite{Saito20} and \citeauthor{Morik2020}~\cite{Morik2020}, which incorporated user feedback in the learning process, and could dynamically adjust to the changing utility with fairness constraints. However, they focused on the changing utility of items and did not consider the scenario where group labels could be dynamic due to the nature of recommendations being an interactive process. To the best of our knowledge, we make the first attempt on dynamic group fairness, focusing on the changing group labels of items.

\subsection{RL for Recommendation} In order to capture the interactive nature of recommendation scenarios, reinforcement learning (RL) based solutions have become an important topic recently. 
A group of work \cite{li2010contextual,bouneffouf2012contextual,zeng2016online} model the problem as contextual multi-armed bandits, which can easily incorporate collaborative filtering methods \cite{cesa2013gang,zhao2013interactive}.
In the meantime, some literature \cite{shani2005mdp,mahmood2007learning,mahmood2009improving,zheng2018drn,xian2020cafe,xian2019reinforcement} found that it is natural to model the recommendation process as a Markov Decision Process (MDP).
In general, this direction can be further categorized as either \textit{policy-based} \cite{Dulac-ArnoldESC15, zhao2018deep, chen2019large, chen2019top} or \textit{value-based} \cite{zhao2018recommendations, zheng2018drn,pei2019value} methods. 
% While policy-based methods aim to learn a policy that generates an action (e.g. recommended items) based on a state, value-based approaches finds the action with the best estimated quality.
Typically, policy-based methods aim to learn a policy that generates an action (e.g. recommended items) based on a state.
Such policy is optimized through policy gradient and can be either deterministic \cite{Dulac-ArnoldESC15, Silver2014DeterministicPG, Lillicrap2016DDPG, zhao2018deep} or stochastic \cite{chen2019large, chen2019top}.
% Early attempts \cite{Dulac-ArnoldESC15} of policy-based approaches apply deterministic policies \cite{Silver2014DeterministicPG, Lillicrap2016DDPG} that directly construct actions and propose to use continuous item embeddings to model the large action space. 
% And \citeauthor{zhao2018deep} \cite{zhao2018deep} employ a deep Deterministic Policy Gradient framework (DDPG) \cite{Lillicrap2016DDPG} for page-wise recommendation.
% As an alternative track, recently studies \cite{chen2019large, chen2019top} explore the stochastic policies that model a distribution of actions.
% \citeauthor{chen2019large} \cite{chen2019large} utilize a balanced hierarchical clustering tree to model the item distribution.
% \citeauthor{chen2019top} \cite{chen2019top} uses an off-policy correction framework to address the data bias issue in top-K recommender system.
On the other hand, value-based methods aims to model the quality (i.e. Q-value) of actions so that the best action corresponds to the one with best value.
% \citeauthor{zhao2018recommendations} \cite{zhao2018recommendations} proposes to use Deep Q-Network (DQN) to estimate the Q-value of a state-action pair in RL and incorporate both positive and negative feedback to learn optimal recommendation strategies.
% And \citeauthor{zheng2018drn} \cite{zheng2018drn} uses a dueling Q-Network to model the Q-value.
% \citeauthor{liu2020end} \cite{liu2020end} later finds that the embedding component of RL cannot always be nicely trained thus proposed an End-to-End framework with additional supervised learning signal.
% To our knowledge, neither policy-based nor value-based methods have discussed how to integrate fairness into the recommendation process.

There also exist several works considering using RL to solve fairness problems in machine learning \cite{Wen2019FairnessWD, jabbari2017fairness}. 
\citeauthor{jabbari2017fairness} \cite{jabbari2017fairness} considered to optimize the 
% aimed to solve a specific fairness modified the
meritocratic fairness defined in \cite{joseph2016fairness} based on long-term rewards.
Their work is designed for a specific fairness constraint and is not suitable for our problem setting.
\citeauthor{Wen2019FairnessWD} \cite{Wen2019FairnessWD} studied a reinforcement learning problem under group fairness constraint, where the state consists of both the feature and the sensitive attributes. 
They developed model-free and model-based methods to learn a decision rule to achieve both demographic parity and near-optimal fairness. 
Different from our work that focuses on item-side fairness, they focused on the user-side fairness.

\section{Preliminary}
\subsubsection{\textbf{Markov Decision Processes.}} In this paper, we study reinforcement learning in Markov Decision Processes (MDPs). 
An MDP is a tuple $M = (\mathcal{S}, \mathcal{A}, \mathcal{P}, \mathcal{R}, \mu, \gamma)$, where $S$ is a set of $n$ states,  $\mathcal{A}$ is a set of $m$ actions,
$\mathcal{P}: \mathcal{S} \times \mathcal{A} \times \mathcal{S} \rightarrow [0,1]$ denotes the transition probability function, $\mathcal{R}: \mathcal{S} \times \mathcal{A} \times \mathcal{S} \rightarrow \mathbb{R}$ is the reward function, $\mu: \mathcal{S} \rightarrow [0,1]$ is the starting state distribution, and $\gamma \in [0,1)$ is the discount factor. 
A stationary policy $\pi: \mathcal{S} \rightarrow P(\mathcal{A})$ is a map from states to probability distributions over actions, with $\pi(a | s)$ denoting the probability of selecting action $a$ in state $s$. 
We denote the set of all stationary policies by $\Pi$.
% In reinforcement learning, we aim to learn a policy $\pi$, which is able to maximize a performance measure, $J(\pi)$, which is typically taken to be the infinite horizon discounted total return, 
In reinforcement learning, we aim to learn a policy $\pi$, which maximizes the infinite horizon discounted total return $J(\pi)$, 
\begin{equation}
\label{eq:discounted_return}
\small
    J(\pi) \doteq \underset{\tau \sim \pi}{\mathrm{E}}\left[\sum_{t=0}^{\infty} \gamma^{\top} R\left(s_{t}, a_{t}, s_{t+1}\right)\right],
\end{equation}
where $\tau$ denotes a trajectory, i.e., $\tau=(s_{0}, a_{0}, s_{1}, a_{1}, \dots)$, and $\tau \sim \pi$ is a shorthand indicating that the distribution over trajectories depends on $\pi$ :
$s_{0} \sim \mu, a_{t} \sim \pi\left(\cdot | s_{t}\right), s_{t+1} \sim P\left(\cdot | s_{t}, a_{t}\right)$.
Let $R(\tau)$ denote the discounted return of a trajectory, we express the on-policy value function as $V^{\pi}(s) \doteq$ $\mathrm{E}_{\tau \sim \pi}\left[R(\tau) | s_{0}=s\right]$, the on-policy action-value function as $Q^{\pi}(s, a) \doteq \mathrm{E}_{\tau \sim \pi}\left[R(\tau) | s_{0}=s, a_{0}=a\right]$, and the advantage function as  $A^{\pi}(s, a) \doteq Q^{\pi}(s, a)-V^{\pi}(s)$.

% The discounted state and state-action value functions are denoted by $V_{\pi}$ and $Q_{\pi}$.
% \begin{equation}
% \begin{aligned}
%     V_{\pi}(s) &= E_{\pi}[U_{t}|S_{t}=s]\\
%     Q_{\pi}(s,a) &= E_{\pi}[U_{t}|S_{t}=s,A_{t}=a],
% \end{aligned}
% \end{equation}
% where $U_{t}$ is the discounted cumulative rewards at time $t$ with state $s$,
% \begin{equation}
%      U_{t} = \sum_{i = t+1}^{\top} \gamma^{i-t-1}R_{i}.
% \end{equation}

% Thus, the optimal state and state-action value function can be defined as,
% \begin{equation}
% \begin{aligned}
%     V_{\pi}^{*}(s) &= \max_{\pi} V_{\pi}(s)\\
%     Q_{\pi}^{*}(s,a) &=  \max_{\pi} Q_{\pi}(s,a).
% \end{aligned}
% \end{equation}

\subsubsection{\textbf{Constrained Markov Decision Processes.}} 
A Constrained Markov Decision Process (CMDP) is an MDP augmented with constraints that restrict the set of allowable policies for that MDP. 
In particular, the MDP can be constrained with a set of auxiliary cost functions $C_{1}, \ldots, C_{m}$ and the corresponding limits $\mathbf{d}_{1}, \ldots, \mathbf{d}_{m}$, which means that the discounted total cost over the cost function $C_{i}$ should be bounded by $\mathbf{d}_i$. Each function $C_{i}: \mathcal{S} \times \mathcal{A} \times \mathcal{S} \rightarrow \mathbb{R}$ maps transition tuples to costs, like the reward in traditional MDP.
Let $J_{C_{i}}(\pi)$ denote the discounted total cost of policy $\pi$ with respect to the cost function $C_{i}$:
\begin{equation}
\small
    J_{C_{i}}(\pi)=\underset{\tau \sim \pi}{\mathrm{E}}\left[\sum_{t=0}^{\infty} \gamma^{\top} C_{i}\left(s_{t}, a_{t}, s_{t+1}\right)\right] .
\end{equation}
The set of feasible stationary policies for a CMDP is then
$
\Pi_{C} \doteq\left\{\pi \in \Pi: \forall i, J_{C_{i}}(\pi) \leq \mathbf{d}_{i}\right\},
$
and the reinforcement learning problem in a CMDP is
$
\pi^{*}=\arg \max _{\pi \in \Pi_{C}} J(\pi).
$, where $J(\pi)$ is the discounted total reward defined in Eq. \eqref{eq:discounted_return}.
% The choice of optimizing only over stationary policies is justified: it has been shown that the set of all optimal policies for a CMDP includes stationary policies, under mild technical conditions. 
% For a thorough review of CMDPs and CMDP theory, we refer the reader to (Altman, 1999 ).
% We refer to $J_{C_{i}}$ as a constraint return, or $C_{i}$-return for short. 
% Finally, we define on-policy value functions, action-value functions, and advantage functions for the auxiliary costs 
Finally, in analogy to $V^{\pi}, Q^{\pi},$ and $A^{\pi}$, we denote these by $V_{C_{i}}^{\pi}, Q_{C_{i}}^{\pi}$, and $A_{C_{i}}^{\pi}$, which replaces reward function $R$ with cost function $C_{i}$, respectively.

\subsubsection{\textbf{Constrained Policy Optimization.}}\label{sec:CPO}
% In local policy search for CMDPs, we additionally require policy iterates to be feasible for the CMDP, so instead of optimizing over $\Pi_{\theta},$ we optimize over $\Pi_{\theta} \cap \Pi_{C}:$
% $$
% \begin{array}{c}
% \pi_{k+1}=\arg \max _{\pi \in \Pi_{\theta}} J(\pi) \\
% \text { s.t. } J_{C_{i}}(\pi) \leq d_{i} \quad i=1, \ldots, m \\
% D\left(\pi, \pi_{k}\right) \leq \delta
% \end{array}
% $$
Inspired by trust region methods \cite{DBLP:journals/corr/SchulmanLMJA15}, \citeauthor{DBLP:journals/corr/AchiamHTA17} \cite{DBLP:journals/corr/AchiamHTA17} proposed Constrained Policy Optimization (CPO), which uses a trust region instead of penalties on policy divergence to enable larger step sizes.
CPO has policy updates of the following form:
\begin{equation} \label{eq:cpo}
\small
\begin{aligned}
\pi_{k+1}&=\arg \max _{\pi \in \Pi_{\theta}}  \underset{\underset{a \sim \pi} {s \sim d^{\pi_{k}} }}{\mathrm{E}}\left[A^{\pi_{k}}(s, a)\right], \\
\text { s.t. } & J_{C_{i}}\left(\pi_{k}\right)+\frac{1}{1-\gamma} \underset{\underset{a \sim \pi} {s \sim d^{\pi_{k}} }}{\mathrm{E}}\left[A_{C_{i}}^{\pi_{k}}(s, a)\right] \leq \mathbf{d}_{i}, \forall i \\
& \bar{D}_{K L}\left(\pi \| \pi_{k}\right) \leq \delta
\end{aligned}
\end{equation}
where $\Pi_{\theta} \subseteq \Pi$ is a set of parameterized policies with parameters $\theta$ (e.g., neural networks with fixed architecture), $d^{\pi_{k}}$ is the state distribution under policy $\pi_{k}$, $\bar{D}_{K L}$ denotes the average KL-divergence, and $\delta > 0$ is the step size.
The set $\left\{\pi_{\theta} \in \Pi_{\theta}: D_{K L}\left(\pi|| \pi_{k}\right) \leq \delta\right\}$ is called the trust region.
Particularly, for problems with only one linear constraint, there is an analytical solution, which is also given by \citeauthor{DBLP:journals/corr/AchiamHTA17} \cite{DBLP:journals/corr/AchiamHTA17}.
% supplementary material (Theorem 2).
Denoting the gradient of the objective in Eq. \eqref{eq:cpo} as $g$, the gradient of constraint as $b$, the Hessian of the KL-divergence as $H$, and defining $c = J_C(\pi_k) - d$, the approximation to Eq. \eqref{eq:cpo} is
\begin{equation} \label{eq:cpo_approx}
\small
\begin{aligned}
\theta_{k+1} &= \arg \max _{\theta}\  g^{\top}\left(\theta-\theta_{k}\right) \\
\text { s.t. } & c+b^{\top}\left(\theta-\theta_{k}\right) \leq 0 \\
& \frac{1}{2}\left(\theta-\theta_{k}\right)^{\top} H\left(\theta-\theta_{k}\right) \leq \delta
\end{aligned}
\end{equation}

% \begin{equation} \label{eq:direction}
% d\theta^{*}=-\frac{1}{\lambda^{*}} H^{-1}\left(g+\nu^{*} b\right)
% \end{equation}
% where $\lambda^{*}$ and $\nu^{*}$ are defined by

% \begin{equation*}
% % \small
% \begin{array}{l}
% \nu^{*}=\left(\frac{\lambda^{*} c-r}{s}\right)_{+} \\
% \lambda^{*}=\underset{\lambda \geq 0}{\arg \max}\left\{\begin{array}{ll}
%  \frac{1}{2 \lambda}\left(\frac{r^{2}}{s}-q\right)+\frac{\lambda}{2}\left(\frac{c^{2}}{s}-\delta\right)-\frac{r c}{s}\ &\text { if } \lambda c-r>0\\
% -\frac{1}{2}\left(\frac{q}{\lambda}+\lambda \delta\right)\ &\text { otherwise }
% \end{array}\right.
% \end{array}
% \end{equation*}
% with $q=g^{\top}H^{-1}g$, $r=g^{\top}H^{-1}b$, and $s=b^{\top}H^{-1}b$.
% 
A more comprehensive review of CMDPs and CPO can be seen in \cite{altman1999constrained} and \cite{ DBLP:journals/corr/AchiamHTA17} respectively.

\section{Problem Formulation}
In this section, we first describe a CMDP that models the recommendation process with general constraints, and then, we describe several fairness constraints, which are suitable for recommendation scenarios. Finally, we combine these two parts together and introduce the fairness-constrained optimization problem.

\subsection{\textbf{CMDP for Recommendation}} 
In each timestamp ($t_1$, $t_2$, $t_3$, $t_4$, $t_5$, $\dots$), when a user sends a request to the recommendation system, the recommendation agent $G$ will take the feature representation of the current user and item candidates $\mathcal{I}$ as input, and generate a list of items $L\in\mathcal{I}^K$ to recommend, where $K \geq 1$.
User $u$ who has received the list of recommended item/items $L$ will give his/her feedback $B$ by his/her clicks on this set of items.
Thus, the state $s$ can be represented by user features (e.g., user's recent click history), action $a$ is represented by items in $L$, reward $r$ is the immediate reward (e.g., whether user clicks on an item in $L$) by taking action $a$ in the current state, and cost $c$ is the immediate cost (e.g., whether the recommended item/items come from the sensitive group).
\begin{itemize}
    \item \textbf{State $\mathcal{S}$:} A state $s_t$ is the representation of user’s most recent positive interaction history $H_t$ with the recommender, as well as his/her demographic information (if exists).
    
    \item \textbf{Action $\mathcal{A}$:} An action $a_t\ = \{ a_t^{1},\ \dots,\ a_t^{K}\}$ is a recommendation list with $K$ items to a user $u$ at time $t$ with current state $s_t$.
    
    \item \textbf{Reward $\mathcal{R}$:} Given the recommendation based on the action $a_t$ and the user state $s_t$, the user will provide his/her feedback, i.e., click, skip, or purchase, etc. The recommender receives immediate reward $R(s_t, a_t)$ according to the user’s feedback.
    
    \item \textbf{Cost $\mathcal{C}$:} Given the recommendation based on the action $a_t$, the environment provides a cost value based on the problem-specific cost function, i.e., the number of items in the recommendation list that come from the sensitive group, and sends the immediate cost $C(s_t, a_t)$ to the recommender.
    
    \item \textbf{Discount rate $\mathcal{\gamma}_r$ and $\mathcal{\gamma}_c$:} $\mathcal{\gamma}_r \in [0,1]$ is a factor measuring the present value of long-term rewards, while $\mathcal{\gamma}_c \in [0,1]$ is another factor measuring the present value of long-term costs.
    % In the case of $\mathcal{\gamma}$ = 0, the recommender considers only immediate rewards but long-term rewards are ignored. 
    % On the other hand, when $\mathcal{\gamma}$ = 1, the recommender treats immediate rewards and long-term rewards as equally important.
    
\end{itemize}

\subsection{Fairness Constraints}
To be consistent with the previous definition in CMDP for recommendation and solve the dynamic change of underlying labels, we define analogs of several frequently proposed fairness constraints.

\subsubsection{\textbf{Demographic Parity Constraints}}
Following~\cite{singh2018fairness}, we can use exposure to define the fairness between different groups of items. 
Demographic parity requires that the average exposure of the items from each group is equal. 
In our setting, we enforce this constraint at each iteration $t$. 
Denoting the number of exposure in a group at iteration $t$ as
\begin{equation} \label{eq:expo}
\small
    \text {Exposure}_t \left(G_{j} \right) = \underset{a^l_t \in a_t}{\sum} \vmathbb{1} (a^l_t \in G_{j}),\ l=1,...,K\\
\end{equation}
% \subsubsection{Demographic Parity Constraints}
% \begin{equation}
% \frac{\mathbb{E}_{(s, a) \sim \Lambda_{\operatorname{maj}}^{(\pi)}} \left[\rho_{s, a}\right]}{U(G_{maj})} = \frac{\mathbb{E}_{(s, a) \sim \Lambda_{\min }^{(\pi)}} \left[\rho_{s, a}\right]}{U(G_{min})}
% \end{equation}
% \noindent
Then we can express demographic parity constraint as follows,
\begin{equation}
\small
    \frac{\text {Exposure}_t\left(G_{0}\right)}{|G_{0}|}=\frac{\text {Exposure}_t\left(G_{1} \right)}{|G_{1}|},
\end{equation}
where groups $G_0$ and $G_1$ are divided based on the item popularity in the recommendation scenario.
% \begin{equation}
%     Cost_{1} = \big| \text { Exposure}\left(G_{0}\right)|G_{1}| - \text { Exposure}\left(G_{1} \right)|G_{0}|  \big|
% \end{equation}

\subsubsection{\textbf{Exact-$K$ Fairness Constraints}}
% \begin{equation}
%     \text { Exposure }\left(G_{0}\right)=\text { Exposure }\left(G_{1} \right)
% \end{equation}
% Inspired by \cite{zehlike2017fa}, 
We define an Exact-$K$ fairness in ranking that requires the proportion/chance of protected candidates in every recommendation list with length $K$ remains statistically below or indistinguishable from a given maximum $\alpha$. 
This kind of fairness constraint is more suitable and feasible in practice for recommender systems as the system can adjust the value of $\alpha$. 
The concrete form of this fairness is shown as below,
\begin{equation} \label{eq:alpha_expo}
\small
    \frac{\text {Exposure}_t\left(G_{0}\right)}{\text {Exposure}_t\left(G_{1} \right)} \leq \alpha
\end{equation}
Note that when $\alpha=\frac{|G_0|}{|G_1|}$ and the equation holds strictly, the above expression would be exactly the same as demographic parity.
% \begin{equation}
%     \frac{\text {Exposure}_t\left(G_{0}\right)}{\alpha}=\frac{\text {Exposure}_t\left(G_{1} \right)}{1-\alpha}
% \end{equation}

% \begin{equation}
%     Cost_{2} = \big|(1-\alpha) \text {  Exposure }\left(G_{0}\right) - \alpha \text { Exposure }\left(G_{1} \right) \big|
% \end{equation}

\subsection{\mbox{\!\!\!\!FCPO: Fairness Constrained Policy Optimization}}
\begin{figure}[t]
% \vspace{-10pt}
\centering
\mbox{
% \hspace{10pt}
\centering
    \includegraphics[scale=0.36]{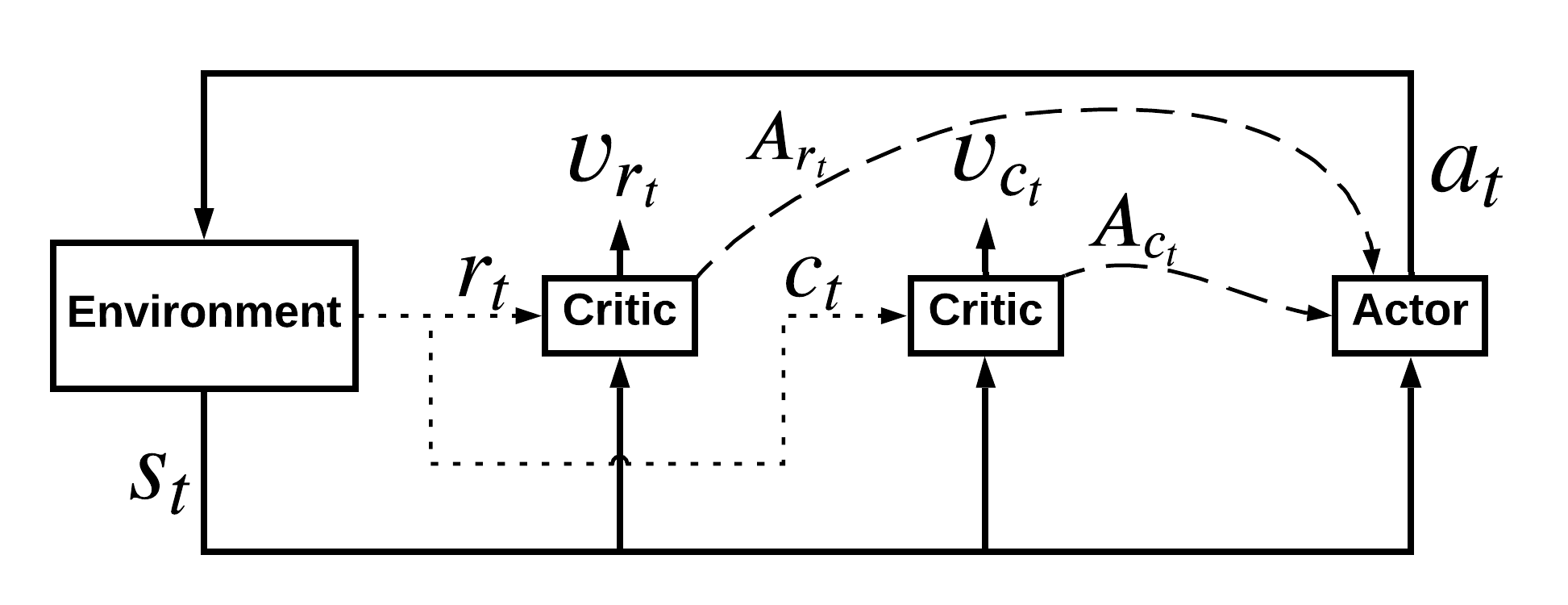}
}
\caption{Illustration of the proposed method.}
\label{fig:model}
\vspace{-15pt}
\end{figure}

% \ykcomment{We can call it FICO?}
% \yzcomment{Maybe FairRL?}
% To overcome the problem of static fairness constraint, we use CMDP to model the problem and use CPO to optimize the problem as CPO is guaranteed for near-constraint satisfaction at each iteration, which means that 
% Once we considered the above group fairness constraints of item exposure in recommendation scenario, 
An illustration of the proposed FCPO is shown in Fig. \ref{fig:model}, containing one actor and two critics. Our goal is to learn the optimal policy for the platform, which is able to maximize the cumulative reward under a certain fairness constraint, as mentioned in previous section.
Specially, in this work, the reward function and the cost function are defined as 
\begin{equation} \label{eq:reward}
\centering
\small
\begin{aligned}
    R(s_{t},a_{t}, s_{t+1}) &= \sum_{l=1}^{K} \vmathbb{1} (a_{t}^l \text{ gets positive feedback})\\
\end{aligned}
\end{equation}

\begin{equation} \label{eq:cost}
\small
\centering
\begin{aligned}
    C(s_{t},a_{t}, s_{t+1}) &=  \sum_{l=1}^{K} \vmathbb{1}(a_{t}^l\ is\ in\ sensitive\ group)
\end{aligned}
\end{equation}
where $a_{t} =  \{ a_{t}^1,\ \dots,\ a_{t}^K \}$ represents a recommendation list including $K$ item IDs, which are selected by the current policy at time point $t$. 
We can see that the expression of cost function is the same as Eq. \eqref{eq:expo}, which represents the total number of items in a specific group exposed to users at time $t$.
Let us consider the sensitive group as group $G_0$, then we have
\begin{equation*} 
\small
\begin{aligned}
    \frac{\text {Exposure}_t\left(G_{0}\right)}{\text {Exposure}_t\left(G_{1} \right)} \leq &\alpha \\
    \text {Exposure}_t\left(G_{0}\right) \leq &\alpha \text {Exposure}_t\left(G_{1} \right)\\
    (1+\alpha)\text {Exposure}_t\left(G_{0}\right) \leq &\alpha \text{Exposure}_t\left(G_{0}\right) + \alpha \text {Exposure}_t\left(G_{1} \right)\\
    (1+\alpha)\text {Exposure}_t\left(G_{0}\right) \leq &\alpha K\\
    C(s_{t},a_{t}, s_{t+1}) \leq &\frac{\alpha}{1+\alpha} K = \alpha^\prime K 
\end{aligned}
\end{equation*}
% Based on the fairness constraints defined previously, we can modify near-constraint satisfaction. 
% In particular, we want to ensure that policy $\pi$ does not favor the $majority\ subpopulation$ over the $minority\ subpopulation$. 
% if the items in the recommendation lists contain a required proportion of members from the protected group, i.e. $G_0$.
Let $C \leq \alpha^{\prime} K$ be satisfied at each iteration, we can get the discounted  total cost,
\begin{equation} \label{eq:fc}
\small
    J_{C}(\pi)=\underset{\tau \sim \pi}{\mathrm{E}}\left[\sum_{t=0}^{T} \gamma^{T}_c\ C\left(s_{t}, a_{t}, s_{t+1}\right)\right] \leq  \sum_{t=0}^T \gamma_{c}^t\ \alpha^{\prime} K
\end{equation}
where $T$ is the length of a recommendation trajectory. 
Eq. \eqref{eq:fc} is the group fairness constraint for our optimization problem and we can denote the limit of the unfairness $\mathbf{d}$ as 
\begin{equation}\label{eq:fairness_limit}
\small
    \mathbf{d} = \sum_{t=1}^T \gamma_{c}^t\ \alpha^{\prime} K.
\end{equation}
Once we finished defining the specific CMDP for recommendation and we have the specific reward function Eq. \eqref{eq:reward}, cost function Eq. \eqref{eq:cost} and the limit of the constraint $\mathbf{d}$, we can take them to Eq. \eqref{eq:cpo} and build our fairness constrained policy optimization framework.
It is worth noting that our model contains only one linear fairness constraint, therefore, as mentioned in Preliminary, we can get an analytical solution by solving Eq. \eqref{eq:cpo_approx} if the problem is feasible. 
We then introduce the framework in the following section.
% to solve it in next section.

% \ykcomment{Is this the problem we aim to solve? If so, what is the difference from original CPO problem? Looks like only the constraints are different? Better to explicitly say we aim to minimize the cost function.}

% We assume the action space has the form $A = \Tilde{A} \times Z$, where $Z = \{min, maj\}$ encodes whether an action $(\Tilde{a},z) \in A$ is from the majority or minority subpopulation, and $\Tilde{A}$ encodes non-sensitive features. 

% \textbf{Definition} Let $M$ be an CMDP with action space of the form $A = \Tilde{A} \times Z$, where $Z = \{maj, min\}$, and let $c \in \mathbb{R}^{\|S\| \times \|A\|}$ be the agent costs. Then, a policy $\pi$ satisfies demographic parity if

% \begin{equation}
% \mathbb{E}_{(s, a) \sim \Lambda_{\operatorname{maj}}^{(\pi)}} \left[c_{s, a}\right] = \mathbb{E}_{(s, a) \sim \Lambda_{\min }^{(\pi)}} \left[c_{s, a}\right],
% \end{equation}
% where \(\Lambda_{z}^{(\pi)}=\Lambda^{(\pi)} | \exists \tilde{a} \in \tilde{A}\).

\section{Proposed Framework}

Our solution to the aforementioned fairness constrained optimization problem follows an Actor-Critic learning scheme, but with an extra critic network designed for the fairness constraint.
In this section, we illustrate how to construct and learn each of these components.

\subsection{The Actor}
\begin{figure}[t]
\vspace{-10pt}
\centering
\mbox{
\hspace{-10pt}
\centering
    \includegraphics[scale=0.34]{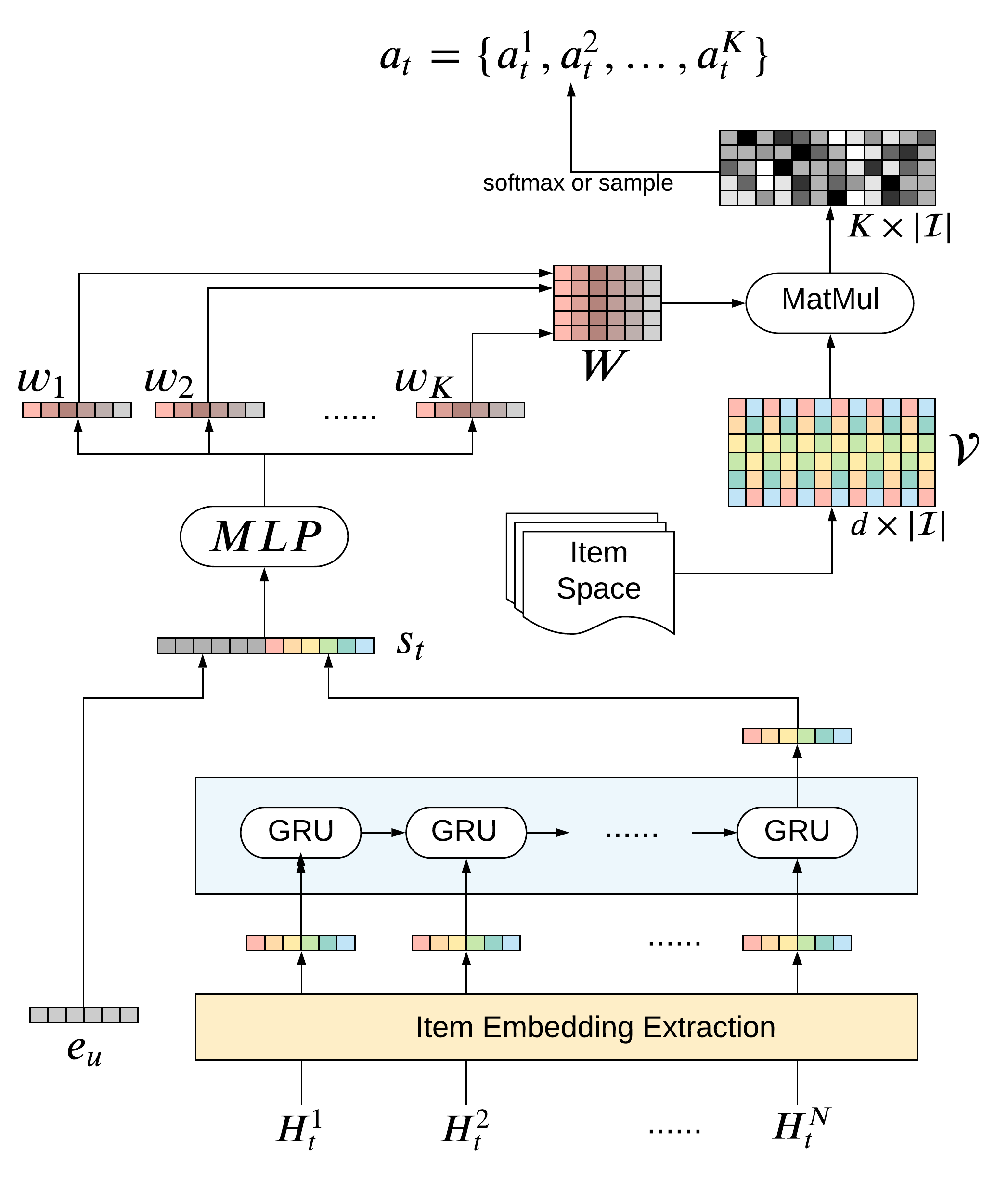}
}
\vspace{-10pt}
\caption{The architecture of the Actor. 
$\theta$ consists of parameters of both the Actor network in $f_\theta$ and the state representation model in Eq. \eqref{eq:state_rep}.}
\label{fig:state}
% \vspace{-15pt}
\end{figure}

The actor component $\pi_\theta$ parameterized by $\theta$ serves as the same functionality as a stochastic policy that samples an action $a_t\in\mathcal{I}^K$ given the current state $s_t\in\mathbb{R}^m$ of a user.
% \begin{equation}
%     a_t \sim \pi_\theta(\cdot|s_t)
% \end{equation}
% \ykcomment{what is the use of $\pi$ here?}
As depicted in Fig. \ref{fig:state}, $s_t$ is first acquired by extracting and concatenating the user embedding $\mathbf{e}_u\in\mathbb{R}^d$ and user's history embedding $\mathbf{h}_u$:
\begin{equation}\label{eq:state_rep}
% s_t = [\bm{v}_u, h(H_t|u)]\label{eq:state_rep}
s_t = [\mathbf{e}_u; \mathbf{h}_u],~\mathbf{h}_u=\mathrm{GRU}(H_t)
\end{equation}
where $H_t = \{H_t^1, H_t^2, \dots, H_t^N\}$ denotes the most recent $N$ items from user $u$'s interaction history, and the history embedding $\mathbf{h}_u$ is acquired by encoding $N$ item embeddings via Gated Recurrent Unites (GRU) \cite{Cho2014gru}.
Note that the user's recent history is organized as a queue, and it is updated only if the recommended item $a_t^l \in a_t$ receives a positive feedback,
\begin{equation}
\small
    H_{t+1}=\left\{
    \begin{array}{cc}
         \{H_t^2,\ \dots,\ H_t^N,\ a_t^l\} & r_t^l>0 \\
         H_t & \text{Otherwise}
    \end{array}
    \right.
\end{equation}
This ensures that the state can always represent the user's most recent interests. 
% \begin{equation}\label{eq:state_prime}
% \small
%     s_{t+1} = [\mathbf{e}_u;\mathrm{GRU}(H_{t+1})]
% \end{equation}

We assume that the probability of actions conditioned on states follows a continuous high-dimensional Gaussian distribution with mean $\mu\in\mathbb{R}^{Kd}$ and covariance matrix $\Sigma\in\mathbb{R}^{Kd\times Kd}$ (only elements at diagonal are non-zeros and there are actually $Kd$ parameters).
For better representation ability, we approximate the distribution via a neural network
% neural network $\pi_\theta: \mathbb{R}^m \rightarrow \mathbb{R}^{K \times d}, \mathbb{R}^{K\times d \times d}$, which 
that maps the encoded state $s_t$ to $\mu$ and $\Sigma$.
% \ykcomment{double check dimensions!!!}
Specifically, we adopt a Multi Layer Perceptron (MLP) with tanh($\cdot$) as the non-linear activation function, i.e. $(\mu,\Sigma)=\mathrm{MLP}(s_t)$.
Then, we can sample a vector from the Gaussian distribution $\mathcal{N}(\mu,\Sigma)$ and convert it into a proposal matrix $W\sim \mathcal{N}(\mu,\Sigma)\in\mathbb{R}^{K\times d}$,
whose $k$-th row, denoted by $W_k\in\mathbb{R}^d$, represents a proposed ``ideal'' item embedding.
% Taking the state representation $s_t$ as input, a mapping function $f_\theta: \mathbb{R}^m \rightarrow \mathbb{R}^{K\times d}$ will generate a proposal matrix $W \sim \mathcal{N}(\mu,\Sigma)$ where $\mu,\Sigma$ are yielded by $f_\theta(s_t)$, each row of which represents a proposed ``ideal'' item embedding.
% \ykcomment{$f$ generates mean and variances of gaussian distributions.}
% And example of such $f$ is a Multi-Layer Perceptron with variation output \cite{}.
% \ykcomment{we shall specifically define how $f$ looks like.}
Then, the probability matrix $P\in\mathbb{R}^{K\times |\mathcal{I}|}$ of selecting the $k$-th candidate item is given by:
\begin{equation}\label{eq:weights}
\small
    P_k = \mathrm{softmax}(W_k \mathcal{V}^\top),~ k=1,\ldots,K,
\end{equation}
where $\mathcal{V}\in\mathbb{R}^{|\mathcal{I}|\times d}$ is the embedding matrix of all candidate items.
This is equivalent to using dot product to determine similarity between $W_k$ and any item.
As the result of taking the action at step $t$, the Actor recommends the $k$-th item as follows:
%or apply sampling during during $\epsilon$-greedy exploration of RL:
\begin{equation} \label{eq:action}
\small
    a_t^k = \argmax_{i\in \{1,\dots,|\mathcal{I}|\}} P_{k,i},~ \forall k=1,\ldots,K,
% \begin{cases}
%     \argmax_i P_{k,i} \forall k\in\{1,2,\dots,K\} & \text{if recommend}\\
%     \mathrm{Bernoulli}_i(P_{k,i}) & \text{if exploration}
% \end{cases}
\end{equation}
where $P_{k,i}$ denotes the probability of taking the $i$-th item at rank $k$.
% is designed to generate a recommendations list according to user's consuming preference.
% As is mentioned in previous section, the user state, denoted by the user's latest clicked/purchased items as well as his/her demographic information, is the input of the policy network.

\subsection{The Critics}
\subsubsection{\textbf{Critic for Value Function}}
% Given the action represented by the actor network discussed in previous section, 
A Critic network $V_\omega(s_t)$ is constructed to approximate the true state value function $V_\omega^\pi(s_t)$ and be used to optimize the actor.
The Critic network is updated according to temporal-difference learning that minimizes the MSE:
\begin{equation} \label{eq:value_update}
\small
    \mathcal{L}(\omega) = \sum_t \Big(y_t - V_\omega(s_t)\Big)^2
\end{equation}
where $y_t = r_t + \gamma_r V_{\omega}(s_{t+1})$.

% Based on the deterministic policy gradient theorem, one can optimize the Actor in the direction of improving the performance of $a$ by following the sampled policy gradient:
% \begin{equation}
%     \bigtriangledown J(\pi_\theta) \approx \frac{1}{N} \sum_{i=1}^N \sum_{t} \bigtriangledown_a V_\omega (s,a)|_{s=s_t,a=\pi_\theta(s_t)} \bigtriangledown_\theta \pi_\theta(s)|_{s=s_t}
% \end{equation}
% where $N$ denotes the mini-batch size.
% On the other hand, 

\subsubsection{\textbf{Critic for Cost Function}}
In addition to the accuracy performance, we introduce a separate Critic network $V_\phi(s)$ for the purpose of constrained policy optimization as explained in section \ref{sec:CPO}, which is updated similarly with Eq. \eqref{eq:value_update}, 
\begin{equation} \label{eq:cost_update}
\small
    \mathcal{L}(\phi) = \sum_t \Big(y_t - V_\phi(s_t)\Big)^2
\end{equation}
where $y_t = c_t + \gamma_c V_{\phi}(s_{t+1})$.
% $V_\phi(s_t)$ is constructed to approximate the true state cost function $V_\phi^\pi(s_t)$ and be used to optimize the actor.
% \begin{equation}
% V_\phi\left(s_{t}\right) \leftarrow V_\phi\left(s_{t}\right)+\alpha\left[c_{t+1}+\gamma_c V_\phi\left(s_{t+1}\right)-V_\phi\left(s_{t}\right)\right]\label{eq:cost_update}
% \end{equation}

\subsection{Training Procedure}
We also present the detailed training procedure of our model in Algorithm \ref{alg:FCPO}. 
In each round, there are two phases --- the trajectory generation phase (line 4-13) and model updating phase (line 14-23), where each trajectory contains $T$ transition results between consumer and the recommendation agent.

\subsection{Testing Procedure}
After finishing the training procedure, FCPO gets fine-tuned hyper-parameters and well-trained parameters. 
Then we conduct the evaluation of our model on several public real-world datasets.
Since our ultimate goal is to achieve long-term group fairness of item exposure with dynamically changing group labels, we propose both short-term evaluation and long-term evaluation.

\subsubsection{\textbf{Short-term Evaluation}} This follows Algorithm \ref{alg:FCPO}, while the difference from training is that it only contains the trajectory generation phase without any updates to the model parameters. 
Once we receive the recommendation results in all trajectories, namely $a_t$, we can use the log data to calculate the recommendation performance, and compute the fairness performance based on the exposure records with fixed group labels. 
We will introduce how to get the initial group label in the experiment part.

\subsubsection{\textbf{Long-term Evaluation}}\label{sec:long_term_eval} This process follows Algorithm \ref{alg:FCPO}, instead of initializing random model parameters, we set well-trained model parameters into our model in advance.
The model parameters will be updated throughout the testing process so as to model an online learning procedure in practice; meanwhile, the item labels will change dynamically based on the current impression results, which means that the fairness constraint will change through time.
To observe long-term performance, we repeatedly recommend $T$ times, so the total number of recommended items is $TK$.
% Meanwhile, through manually control the length of the recommendation list $K$ and the number of recommendation sessions, we can study the long-term performance of the proposed model.

\begin{algorithm}[t]
\small
    % \SetKwInOut{Input}{Input}
    % \SetKwInOut{Output}{Output}
    \textbf{Input:} step size $\delta$, cost limit value $\mathbf{d}$, and line search ratio $\beta$ \\
    \textbf{Output:} parameters $\theta$, $\omega$ and $\phi$ of actor network, value function, cost function \\
    % \underline{function Euclid} $(a,b)$\;
    % \Input{}
    % \Output{$\pi_\theta$}
    Randomly initialize $\theta$, $\omega$ and $\phi$. \\
    % set an appropriate cost limit value $\mathbf{d}$ and line search rate $\beta$;\\
    Initialize replay buffer $D$;
    
    \For{$Round\ =\ 1\ ...\ M$}{
        Initialize user state $s_0$ from log data\;
        \For{$t\ =\ 1\ ...\ T$}{
        Observe current state $s_t$ based on Eq. \eqref{eq:state_rep};\\
        Select an action $a_t\ = \{ a_t^{1},\ \dots,\ a_t^{K}\} \in \mathcal{I}^K$ based on Eq. \eqref{eq:weights} and Eq. \eqref{eq:action}\\
        Calculate reward $r_t$ and cost $c_t$ according to environment feedback based on Eq. \eqref{eq:reward} and Eq. \eqref{eq:cost};\\
        Update $s_{t+1}$ based on Eq. \eqref{eq:state_prime};\\
        Store transition $(s_t,a_t,r_t,c_t,s_{t+1})$ in $D$ in its corresponding trajectory.
        }
        
        Sample minibatch of $\mathcal{N}$ trajectories $\mathcal{T}$ from $D$;\\
        Calculate advantage value $A$, advantage cost value $A_c$;\\
        % , discounted cumulative reward, and discounted cumulative cost for each trajectory;\\
        Obtain gradient direction $d_\theta$ by solving Eq. \eqref{eq:cpo_approx} with $A$ and $A_c$;\\
        % Use conjugate gradient method to find update gradient $d_\theta = \sqrt{\frac{2\delta}{\bm{x}^\top\bm{F}\bm{x}}}\bm{x}$\;
        \Repeat{$\pi_{\theta'}(s)$ in trust region \& loss improves
        \& cost $\leq \mathbf{d}$ 
        }{
            $\theta' \leftarrow \theta + d_\theta$\\
            $d_\theta \leftarrow \beta d_\theta$
        }
        
        (Policy update) $\theta \leftarrow \theta'$\;
        (Value update) Optimize $\bm{\omega}$ 
        based on Eq.\eqref{eq:value_update}\;
        (Cost update) Optimize $\bm{\phi}$ 
        based on Eq.\eqref{eq:cost_update}\;
    }
    \caption{Parameters Training for FCPO}
    \label{alg:FCPO}
\end{algorithm}

% \ykcomment{consider to put part of CPO algorithms here to make it like a complete method.}

\section{Experiments}
\subsection{Dataset Description}
We use the user transaction data from $Movielens$  \cite{Harper:2015:MDH:2866565.2827872} in our experiments to verify the recommendation performance of \textbf{FCPO}\footnote{https://github.com/TobyGE/FCPO}. 
We choose $Movielens100K$ and $Movielens1M$ \footnote{\url{https://grouplens.org/datasets/Movielens/}} datasets, which include one hundred thousand and one million user transactions, respectively (user id, item id, rating, timestamp, etc.).

For each dataset, we sort the transactions of each user according to the timestamp, and then split the records into training and testing sets chronologically by 4:1, and the last item of each user in the training set is put into the validation set.
Some basic statistics of the experimental datasets are shown in Table \ref{tab:dataset}. 
% We plot the results of item exposure in Movielen100K and Movielens1M in Fig. \ref{fig:item_expo}.
We split items into two groups $G_0$ and $G_1$ based on item popularity, i.e., the number of exposures for each item. Specifically, the top 20\% items in terms of number of impressions belong to the popular group $G_0$, and the remaining 80\% belong to the long-tail group $G_1$.

Moreover, for RL-based recommenders, the initial state for each user during training is the first five clicked items in the training set, and the initial state during testing is the last five clicked items in the training set. 
For simplicity, each time the RL agent recommends one item to the user, while we can adjust the length of the recommendation list easily in practice.

% \begin{figure}[t]
% \centering
% \mbox{
% \hspace{15pt}
% \centering
%     \includegraphics[scale=0.5]{fig/item_exposure.png}
% }
% \caption{Item exposure in ML-100K and ML-1M}
% \label{fig:item_expo}
% \vspace{-10pt}
% \end{figure}

\subsection{Experimental Setup}\label{sec:experimental_setup}
\textbf{Baselines:} 
We compare our model with the following baselines, including both traditional and RL based methods. 
% For economic methods, we involve baselines that do not consider risk preferences to illustrate the importance of risk consideration.

\begin{itemize}
\item {\bf MF}: Collaborative Filtering based on matrix factorization \cite{koren2009matrix} is a representative method for rating prediction. 
Basically, the user and item rating vectors are considered as the representation vector for each user and item.
% In this experiment, we use CF based on Singular Value Decomposition techniques \cite{koren2009matrix, ricci2011introduction}. 

\item {\bf BPR-MF}: Bayesian Personalized Ranking \cite{bpr} is one of the most widely used ranking methods for top-K recommendation, which models recommendation as a pair-wise ranking problem.
% It considers a pair-wise ranking with two classes that bought and not bought. Instead of classifying with the prediction of a single good, BPR tried to use the difference between two predictions. 
% In the implementation, we conduct balanced negative sampling on unpurchased items for model learning.

\item {\bf NCF}: Neural Collaborative Filtering \cite{he2017neural} is a simple neural network-based recommendation algorithm.
% which is based on deep neural networks. 
In particular, we choose Neural Matrix Factorization to conduct the experiments, fusing both Generalized Matrix Factorization (GMF) and Multiple Layer Perceptron (MLP) under the NCF framework.

\item {\bf LIRD}: It is the short for List-wise Recommendation based on Deep reinforcement learning \cite{DBLP:journals/corr/abs-1801-00209}. The original paper simply utilizes the concatenation of item embeddings to represent the user state. For fair comparison, we replace the state representation with the same structure of FCPO, as is shown is Fig. \ref{fig:state}.
\end{itemize}

In this work, we also include a classical fairness baseline called Fairness Of Exposure in Ranking (FOE) \cite{singh2018fairness} in our experiment to compare the fairness performance with our model.
\textbf{FOE} can be seen as a reranking framework based on group fairness constraints,
% \textbf{FOE} is a computational framework allowing group fairness constraints on ranking in terms of exposure allocation. 
% The authors expressed the problem of finding the utility-maximizing probabilistic ranking under fairness constraint as a linear program, which can be solved with standard algorithms.
and it is originally designed for searching problems, so we made a few modification to accommodate the recommendation task.
% Since FOE can be seen as a reranking framework, it needs a utility measure to get the expected utility of a document $d$ for query $q$. 
We use ranking prediction model such as MF, BPR, and NCF as the base ranker, where the raw utility is given by the predicted probability of user $i$ clicking item $j$. 
In our experiment, we have \textbf{MF-FOE}, \textbf{BPR-FOE} and \textbf{NCF-FOE} as our fairness baselines.
% The reason why there does not exist a \textbf{LIRD-FOE} is that \textbf{LIRD} is a sequence model.
Since FOE assumes independence of items in the list, it cannot be applied to LIRD, which is a sequential model and the order in its recommendation makes a difference.
% We cannot simply rerank the recommendation lists as the future states and actions will be influenced by current actions causally.
Meanwhile, FOE for personalized recommendation needs to solve a linear program with size $|\mathcal{I}| \times |\mathcal{I}|$ for each consumer, which brings huge computational costs.
In order to make the problem feasible, we let FOE rerank top-200 items from the base ranker (e.g. MF), and select the new top-K (K<200) as the final recommendation results.

\begin{table}[t]
\vspace{-10pt}
\caption{Basic statistics of the experimental datasets.}
\label{tab:dataset}
\centering
\setlength{\tabcolsep}{5pt}
\begin{adjustbox}{max width=\linewidth}
\begin{tabular}
    {lcccccccc} \toprule
    Dataset & \#users & \#items & \#act./user & \#act./item & \#act. & density \\\midrule
    Movielens100K & 943 & 1682 & 106 & 59.45 & 100,000 & 6.305\%\\
    Movielens1M & 6040 & 3706 & 166 & 270 &1,000,209 & 4.468\%\\\bottomrule
\end{tabular}
\end{adjustbox}
\vspace{-15pt}
\end{table}

We implement MF, BPR-MF, NCF, MF-FOE, BPR-FOE and NCF-FOE using \textit{Pytorch} with Adam optimizer.
For all the methods, we consider latent dimensions $d$ from \{16, 32, 64, 128, 256\}, learning rate $lr$ from \{1e-1, 5e-2, 1e-2, \dots, 5e-4, 1e-4\}, and the L2 penalty is chosen from \{0.01, 0.1, 1\}. 
We tune the hyper-parameters using the validation set and terminate training when the performance on the validation set does not change within 5 epochs.
% Meanwhile, to avoid heavy computation on all testing user-item pairs, we followed the mechanism in \cite{he2017neural, elkahky2015multi, koren2008factorization}. 
% For each user $i$, we randomly sample 100 negative items (items that are not interacted by the user) and rank these items with the positive sample in the test set. 

We implement \textbf{FCPO} with $Pytorch$ as well. 
We perform PMF \cite{mnih2008probabilistic} to pretrain 100-dimensional user and item embeddings, and fix them through the whole experiment.
We set $|H_t|=5$, and use 2 layer of GRU to get state representation $s_t$. 
For the policy network and each of the two critic networks, we use two hidden layer MLP with tanh($\cdot$) as activation function.
% After that, as described in previous section, we pass user embedding and user's corresponding latest 5 positive feedback $H_t$ through two GRU layers to get state representation $s_t$, and then pass the $s_t$ through two hidden layer with activation function set as Tanh in the policy network.
% For the two Critics, they are designed with the same architecture --- two hidden layer with activation function set as Tanh, 
Critics are learned through LBFGS optimizer \cite{andrew2007scalable}.
Finally, we fine-tune FCPO's hyperparameters on our validation set.
In order to examine the trade-off between performance and fairness, we set different level of fairness constraint controlled by the values of $\alpha^{\prime}$ in Eq.
\eqref{eq:fc} and calculate the limit $\mathbf{d}$ using Eq. \eqref{eq:fairness_limit}.
We denote the resulting alternatives as \textbf{FCPO-1}, \textbf{FCPO-2}, and \textbf{FCPO-3}, whose corresponding fairness be constrained by setting $\alpha^{\prime}=1$, 
% , which is medium fairness constrained by setting 
$\alpha^{\prime}=0.8$, and 
% \textbf{FCPO-3}, which is strong fairness constrained by setting 
$\alpha^{\prime}=0.4$ correspondingly in our experiments.

\textbf{Evaluation Metrics:} 
% Based on the ranking results of all items, 
We adopt several common top-K ranking metrics including \text{Recall}, \text{F1 Score}, and \text{NDCG} to evaluate each model's recommendation performance.
% $F1\ Score$ considers both the precision and the recall of the test to compute the score, while $NDCG$ is a position-aware metric, involving a discount function over the rank.       % \cite{wang2013theoretical}.
In addition to these accuracy-based metrics, we also include two fairness measures -- \text{Gini Index} and \text{Popularity Rate}, with respect to item exposures for individual items and groups, respectively.
Gini Index measures the inequality among values of a frequency distribution (for example, numbers of impressions), which can be seen as an individual level measure.
Given a list of impressions from all items, $\mathcal{M}=[g_{1},g_{2},...,g_{|\mathcal{I}|}]$, the Gini Index can be calculated by Eq.\eqref{eq:gini},
\begin{equation}\label{eq:gini}
\small
    Gini\ Index(\mathcal{G}) = \frac{1}{2|\mathcal{I}|^2\bar{g}}\sum_{i=1}^{|\mathcal{I}|}\sum_{j=1}^{|\mathcal{I}|} |g_{i} - g_{j}|,
\end{equation}
where $\bar{g}$ represents the mean of all item impressions.
Popularity Rate, on the other hand, simply refers to the proportion of popular items in the recommendation list against the total number of items in the list, which can be seen as a popularity level measure of fairness.
Both of the two fairness measures are the smaller, the fairer to the recommender system.

% , and the detailed settings on each dataset for it are shown in Table \ref{tab:parameters}.

\begin{table*}[]
\caption{Summary of the performance on two datasets. 
We evaluate for ranking ($Recall$, $F_1$ and $NDCG$, in percentage (\%) values, \% symbol is omitted in the table for clarity) and fairness ($Gini$ $Index$ and $Popularity$ $Rate$, also in \% values), whiles $K$ is the length of recommendation list. 
When FCPO is the best, its improvements against the best baseline are significant at p < 0.01.}
\centering
\begin{adjustbox}{max width=\linewidth}
\setlength{\tabcolsep}{7pt}
\begin{tabular}
    {m{1.33cm} ccc ccc ccc ccc ccc} \toprule
    \multirow{2}{*}{Methods} 
    % & Metrics 
    & \multicolumn{3}{c}{Recall (\%) $\uparrow$} 
    & \multicolumn{3}{c}{F1 (\%) $\uparrow$} 
    & \multicolumn{3}{c}{NDCG (\%) $\uparrow$} 
    & \multicolumn{3}{c}{Gini Index (\%) $\downarrow$} 
    & \multicolumn{3}{c}{Popularity Rate (\%) $\uparrow$}\\\cmidrule(lr){2-4} \cmidrule(lr){5-7} \cmidrule(lr){8-10} \cmidrule(lr){11-13} \cmidrule(lr){14-16}
 & K=5 & K=10 & K=20 & K=5 & K=10 & K=20 & K=5 & K=10 & K=20 & K=5 & K=10 & K=20 & K=5 & K=10 & K=20 \\\midrule 
 \multicolumn{16}{c}{Movielens-100K} \\\midrule
MF  &  1.847 & 3.785 & 7.443 & 2.457 & 3.780 & 5.074 & 3.591 & 4.240 & 5.684 & 98.99 & 98.37 & 97.03 & 99.98 & 99.96 & 99.92\\
BPR-MF  &  1.304 & 3.539 & 8.093 & 1.824 & 3.592 & 5.409 & 3.025 & 3.946 & 5.787 & 98.74 & 98.17 & 97.01 & 99.87 & 99.87 & 99.78\\
NCF  &  \underline{1.995} & 3.831 & 6.983 & \underline{2.846} & \underline{4.267} & \underline{5.383} & \underline{5.319} & \underline{5.660} & \underline{6.510} & 99.70 & 99.39 & 98.80 & 100.0 & 100.0 & 100.0\\
LIRD  & 1.769 & \underline{5.467} & \underline{8.999} & 2.199 & 4.259 & 4.934 & 3.025 & 3.946 & 5.787 & 99.70 & 99.41 & 98.81 & 100.0 & 100.0 & 100.0\\\midrule

MF-FOE & 1.164 & 2.247 & 4.179 & 1.739 & 2.730 & 3.794 & 3.520 & 3.796 & 4.367 & \underline{86.29} & \underline{84.05} & \underline{82.98} & 92.90 & 91.89 & 90.98 \\
BPR-FOE & 0.974 & 2.053 & 4.404 & 1.496 & 2.568 & 3.933 & 3.127 & 3.514 & 4.332 & 86.50 & 84.38 & 83.78 & \underline{92.17} & \underline{91.36} & \underline{90.70}\\
NCF-FOE & 1.193 & 1.987 & 4.251 & 1.759 & 2.398 & 3.698 & 4.033 & 3.897 & 4.633 & 96.92 & 94.53 & 90.44 & 100.0 & 100.0 & 100.0\\\midrule
% LIRD-FOE  &  6.783 & \underline{5.566} & \underline{4.129} & 15.94 & 19.45 & \underline{22.74}\\\midrule

FCPO-1  &  \textbf{4.740} & \textbf{8.607} & \textbf{14.48} & \textbf{4.547} & \textbf{5.499} & \textbf{5.855} & \textbf{6.031} & \textbf{7.329} & \textbf{9.323} & 98.73 & 98.07 & 96.75 & 92.60 & 90.42 & 85.85\\
FCPO-2  &  3.085 & 5.811 & 10.41 & 3.270 & 4.164 & 4.953 & 4.296 & 5.203 & 7.104 & 97.95 & 96.88 & 94.78 & 70.07 & 68.28 & 65.55\\
FCPO-3  &  0.920 & 1.668 & 3.329 & 1.272 & 1.807 & 2.535 & 2.255 & 2.369 & 2.871 & \textbf{75.23} & \textbf{74.06} & \textbf{73.23} & \textbf{36.52} & \textbf{36.66} & \textbf{36.94}\\\midrule

\multicolumn{16}{c}{Movielens-1M} \\\midrule
MF  &  1.152 & 2.352 & 4.650 & 1.701 & 2.814 & 4.103 & 3.240 & 3.686 & 4.574 &  99.44 & 99.18 & 98.74 & 99.92 & 99.90 & 99.86\\
BPR-MF  &   1.240 & 2.627 & 5.143 & 1.773 & 2.943 & 4.197 & 3.078 & 3.593 & 4.632 & 98.93 & 98.44 & 97.61 & 99.40 & 99.23 & 98.96\\
NCF & 1.178 & 2.313 & 4.589 & 1.832 & 2.976 & \underline{4.382} & \underline{4.114} & \underline{4.380} & \underline{5.080} & 99.85 & 99.71 & 99.42 & 100.0 & 100.0 & 100.0\\
LIRD & \underline{1.961} & \underline{3.656} & \underline{5.643} & \underline{2.673} & \underline{3.758} & 4.065 & 3.078 & 3.593 & 4.632 & 99.87 & 99.73 & 99.46 & 100.0 & 100.0 & 95.00\\\midrule

MF-FOE & 0.768 & 1.534 & 3.220 & 1.246 & 2.107 & 3.345 & 3.321 & 3.487 & 4.021 & 92.50 & 91.06 & 91.32 & 98.89 & 98.78 & 98.68 \\
BPR-FOE  &  0.860 & 1.637 & 3.387 & 1.374 & 2.233 & 3.501 & 3.389 & 3.594 & 4.158 & \underline{90.48} & \underline{88.92} & \underline{89.01} & \underline{96.56} & \underline{96.12} & \underline{95.78}\\
NCF-FOE & 0.748 & 1.403 & 2.954 & 1.230 & 1.980 & 3.175 & 3.567 & 3.589 & 4.011 & 97.73 & 96.57 & 95.04 & 100.0 & 100.0 & 100.0\\\midrule
% LIRD-FOE  &  6.783 & \underline{5.566} & \underline{4.129} & 15.94 & 19.45 & \underline{22.74}\\\midrule

FCPO-1 & \textbf{2.033} & \textbf{4.498} & \textbf{8.027} & \textbf{2.668} & \textbf{4.261} & \textbf{5.201} & \textbf{4.398} & \textbf{5.274} & \textbf{6.432} & 99.81 & 99.67 & 99.34 & 99.28 & 96.93 & 91.70\\
FCPO-2  &  1.520 & 3.218 & 6.417 & 2.015 & 3.057 & 4.145 & 3.483 & 3.920 & 5.133 & 99.47 & 99.10 & 97.41 & 72.66 & 68.27 & 71.35\\
FCPO-3  &  0.998 & 1.925 & 3.716 & 1.449 & 2.185 & 2.948 & 2.795 & 2.987 & 3.515 & \textbf{88.97} & \textbf{88.34} & \textbf{87.70} & \textbf{63.43} & \textbf{62.73} & \textbf{61.45}\\\bottomrule
\end{tabular}\label{tab:result}
\end{adjustbox}
\vspace{-10pt}
\end{table*}

\begin{figure*}[t]
\mbox{
\hspace{-15pt}
\centering
    \subfigure[NDCG vs Negative Gini on ML100K]{\label{fig:ml100k_ndcg_gini}
        \includegraphics[width=0.26\textwidth]{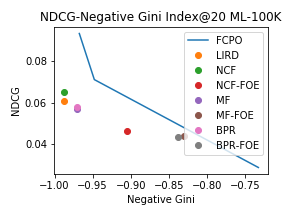}}
    \hspace{-10pt}
    \subfigure[NDCG vs Long-tail Rate on ML100K]{\label{fig:ml100k_ndcg_pop}
        \includegraphics[width=0.26\textwidth]{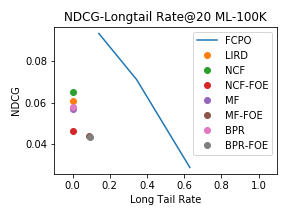}}
    \hspace{-10pt}
    \subfigure[NDCG vs Negative Gini on ML1M]{\label{fig:ml1m_ndcg_gini}
        \includegraphics[width=0.26\textwidth]{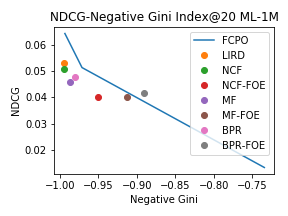}}
    \hspace{-10pt}
    \subfigure[NDCG vs Long-tail Rate on ML1M]{\label{fig:ml1m_ndcg_pop}
        \includegraphics[width=0.26\textwidth]{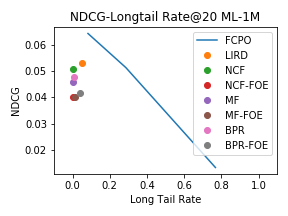}}
}
\caption{NDCG@20 vs. Negative Gini Index@20 and NDCG@20 vs. Long-tail Rate@20 in two datasets. $x$-axis is the negative gini index in \ref{fig:ml100k_ndcg_gini} and \ref{fig:ml1m_ndcg_gini}, and is the long-tail rate in \ref{fig:ml100k_ndcg_pop} and \ref{fig:ml1m_ndcg_pop}; $y$-axis represents the value of NDCG.}
\label{fig:ndcg_fairness}
\vspace{-10pt}
\end{figure*}

\balance
\subsection{Experimental Results}
The major experimental results are shown in Table \ref{tab:result}, besides, we also plot the \textit{NDCG vs. Negative Gini Index} and \textit{NDCG vs. Long-tail Rate} in Fig. \ref{fig:ndcg_fairness} under the length of recommendation list $K=20$. We analyze and discuss the results in terms of the following perspectives.

\subsubsection*{\bf i) Recommendation Performance:} 
For recommendation performance, we compare FCPO-1 with MF, BPR, NCF, and LIRD based on $Recall@k$, $F1@k$ and $NDCG@k$.
The results of the recommendation performance are shown in Table \ref{tab:result}.
The largest value on each dataset and for each evaluation measure is significant at 0.01 level.
Among all the baseline models, NCF is the strongest on Movielens100K: when averaging across recommendation lengths, NCF gets 11.45\% improvement than MF, 18.01\% than BPR, and 6.17 \% than LIRD; and LIRD is the strongest on Movielens1M: when averaging across recommendation lengths, LIRD gets 17.69\% improvement than MF, 14.50\% than BPR, and 9.68 \% than NCF.

Our FCPO approach achieves the best top-K recommendation performance against all baselines on both datasets. 
On the one hand, when averaging across recommendation lengths on Movielens100K, FCPO gets 33.09\% improvement than NCF; on the other hand, when averaging across recommendation lengths on Movielens1M, FCPO gets 18.65 \% improvement than LIRD.
These observations imply that the proposed method does have the ability to capture dynamic user-item interactions, which captures better user preferences resulting in better recommendation results.
Another interesting observation is that FCPO is better than LIRD even though they use the same state representation and similar training procedure. This may be attributed to the trust-region-based optimization method, which stabilizes the model learning process.

% Compared with deep recommendation baseline NCF, our model gets 13.82\% improvement on F1 on average, and especially a 48.52\% improvement for F1@5 on \textit{Ciao} dataset; furthermore, our model also gets 13.46\% improvement than NCF for NDCG on average, and a 43.81\% improvement for NDCG@5 on \textit{Ciao} dataset.
% Meanwhile, in order to prove the validity of our improvement, we also plot the NDCG for different lengths of recommendation lists on all datasets, shown in Fig.\ref{fig:NDCG}. 
% We can find that RARE outperforms all baseline for larger $K$, for example, $K=50$. 
% As shown, the black curve represented as RARE, is always above the best baselines in all six datasets.
% These observations imply that by modeling user behaviors under uncertainty based on established risk-aware principles, our model has the ability to capture better user preferences resulting in better recommendation results.

\subsubsection*{\bf ii) Short-term Fairness Performance:}
For fairness performance, we compare three FCPOs with MF-FOE, BPR-FOE, and NCF-FOE based on $Gini\ Index@k$ and $Popularity\ Rate@k$, which are also shown in Table \ref{tab:result}.
We can easily see that there exists a trade-off between the recommendation performance and the fairness performance both in FCPO and FOE, which is understandable, as most of the long-tail items have relatively fewer user interactions.
In order to better illustrate the trade-off between FCPO and FOE, we fix the length of the recommendation list at 20 and plot NDCG against Negative Gini Index and Long-tail Rate in Fig. \ref{fig:ndcg_fairness} for both datasets, where the long-tail rate is equal to one minus popularity rate.
The blue line represents FCPO under three different levels of fairness constraint.
We choose Negative Gini Index and Long-tail Rate instead of the original ones as they are the bigger, the better, which is easier for comparison.
In most cases, for the same Gini Index, our method achieves much better NDCG; meanwhile, under the same NDCG scores, our method achieves better fairness.
In other words, our method FCPO can achieve much better trade-off than FOE in both individual fairness (measured by Gini Index) and group fairness (measured by Long-tail Rate).
We can see that even with the light fairness constraint, FCPO-1 is better than traditional baselines and the FOE-based methods on group fairness.

% \begin{table}[h]
% \caption{Running time on test set of each method.}
% \label{tab:time}
% \centering
% \setlength{\tabcolsep}{5pt}
% \begin{adjustbox}{max width=\linewidth}
% \begin{tabular}
%     {lcccc} \toprule
%       Running Time (min) &  FOE & FCPO\\\midrule
%       Movielens100K & 943 & 1682 & 106 & 210\\
%       Movielens1M & 943 & 1682 & 106 & 671\\\bottomrule
% \end{tabular}
% \end{adjustbox}
% \end{table}

\subsubsection*{\bf iii) Efficiency Performance:}

We compare FOE-based methods with FCPO in terms of the single-core CPU running time to generate a recommendation list of size $K=100$ for all users.
The running time between the base ranker of FOE-based methods is relatively the same, but the additional reranking step of FOE may take substantial time.
In our observation on Movielens100K dataset, the recommendation time is 90min, 6h30min, and 60h30min for reranking from $200$, $400$, and $800$ items, respectively, while FCPO only takes around 3h and select items from the entire item set (1682 items).
Our observation on Movielens1M dataset shows that FOE-based methods take 10h30min, 43h30min, and 397h to rerank from $200$, $400$, and $800$ items, respectively, while FCPO takes around 11h33min selecting in the entire item set (3706 items).
As mentioned before, these experiments are running on single-core CPU for fair comparison, therefore, we can easily speed them up by using parallel computing. 
% in section \ref{sec:experimental_setup}, FOE becomes impractical when the candidate item sets is large.

\begin{figure}[t]
% \vspace{-6pt}
\mbox{
\hspace{-20pt}
\centering
    \subfigure[NDCG on Movielens100K]{\label{fig:ml100k_long_term_ndcg}
        \includegraphics[width=0.26\textwidth]{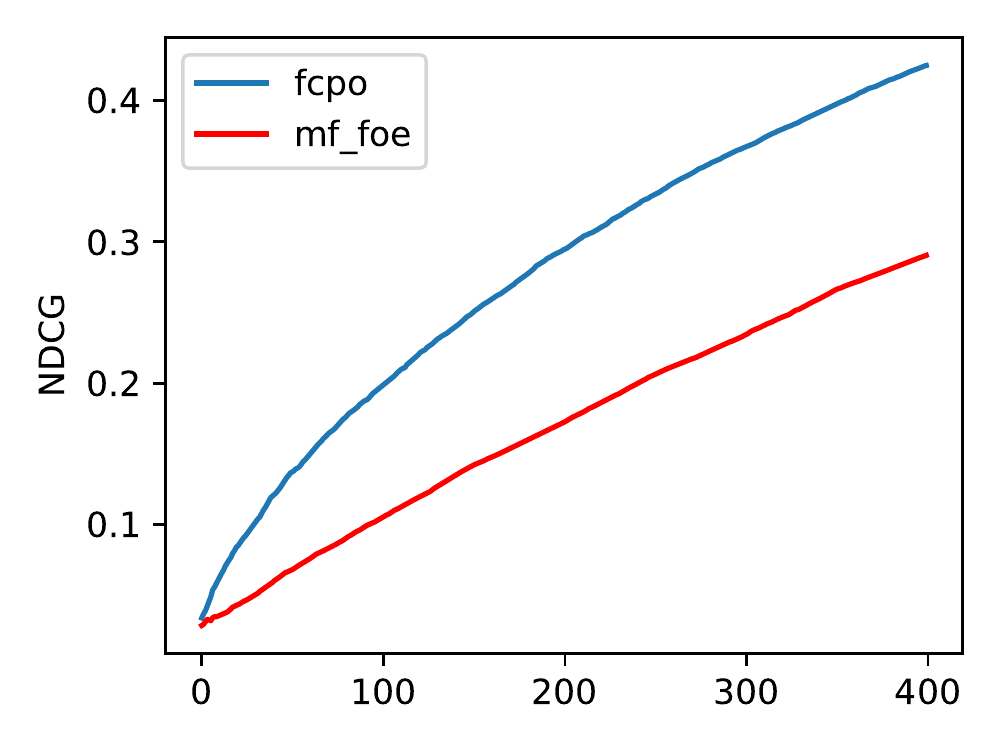}}
    \hspace{-10pt}
    \subfigure[NDCG on Movielens1M]{\label{fig:ml1m_long_term_ndcg}
        \includegraphics[width=0.26\textwidth]{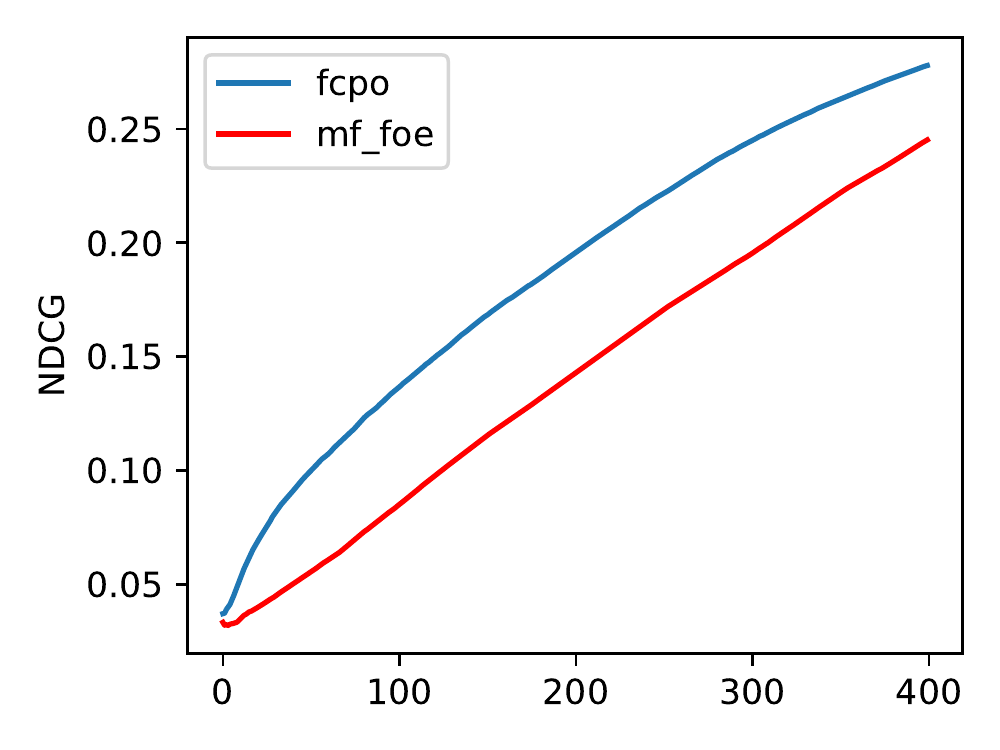}}
}
% \vspace{-10pt}
\mbox{
\hspace{-20pt}
% \vspace{-20pt}
\centering
    \subfigure[Gini on Movielens100K]{
    \label{fig:ml100k_long_term_gini}
        \includegraphics[width=0.26\textwidth]{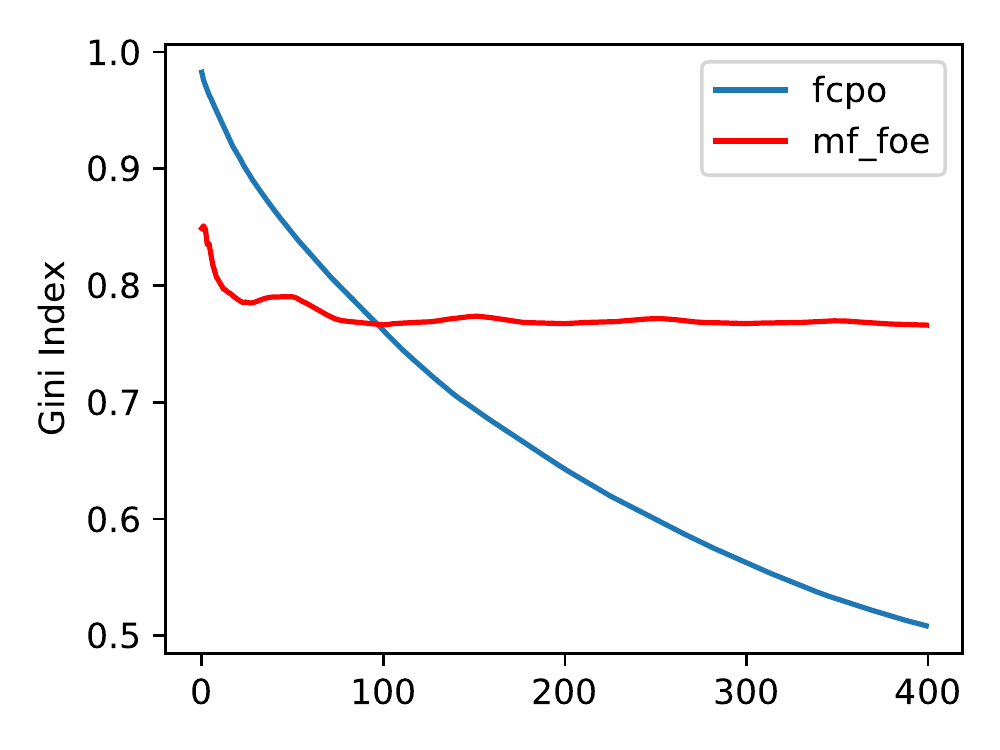}}
    \hspace{-10pt}
    \subfigure[Gini on Movielens1M]{\label{fig:ml1m_long_term_gini}
        \includegraphics[width=0.26\textwidth]{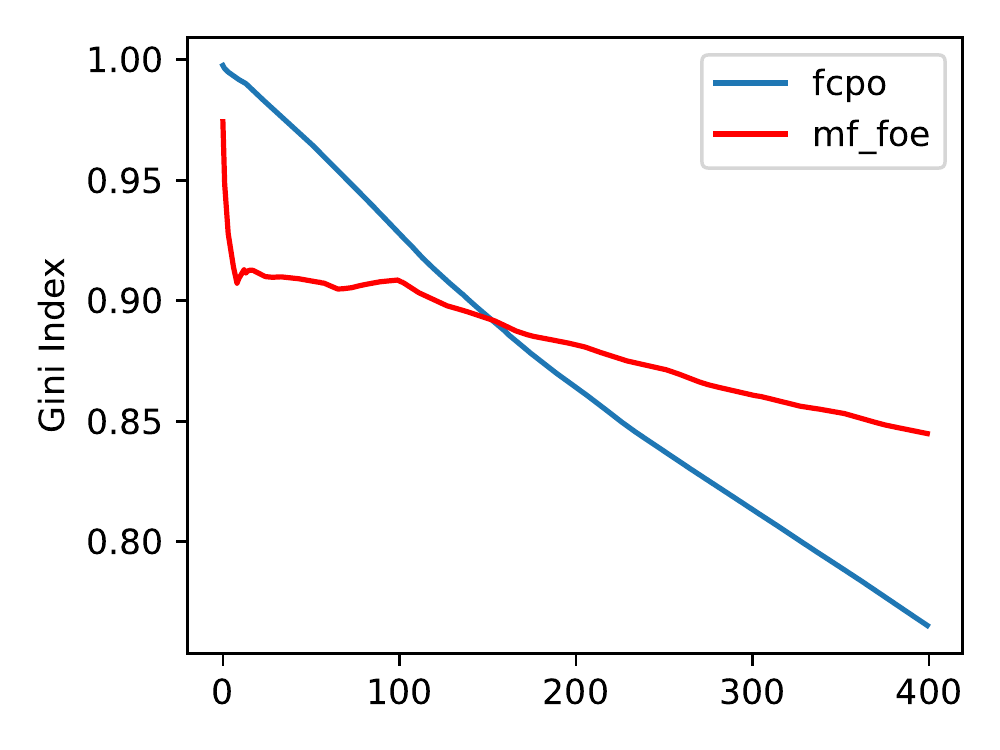}}
}
% \vspace{-10pt}
\mbox{
\hspace{-20pt}
\centering
    \subfigure[Popularity Rate on Movielens100K]{\label{fig:ml100k_long_term_pr}
        \includegraphics[width=0.26\textwidth]{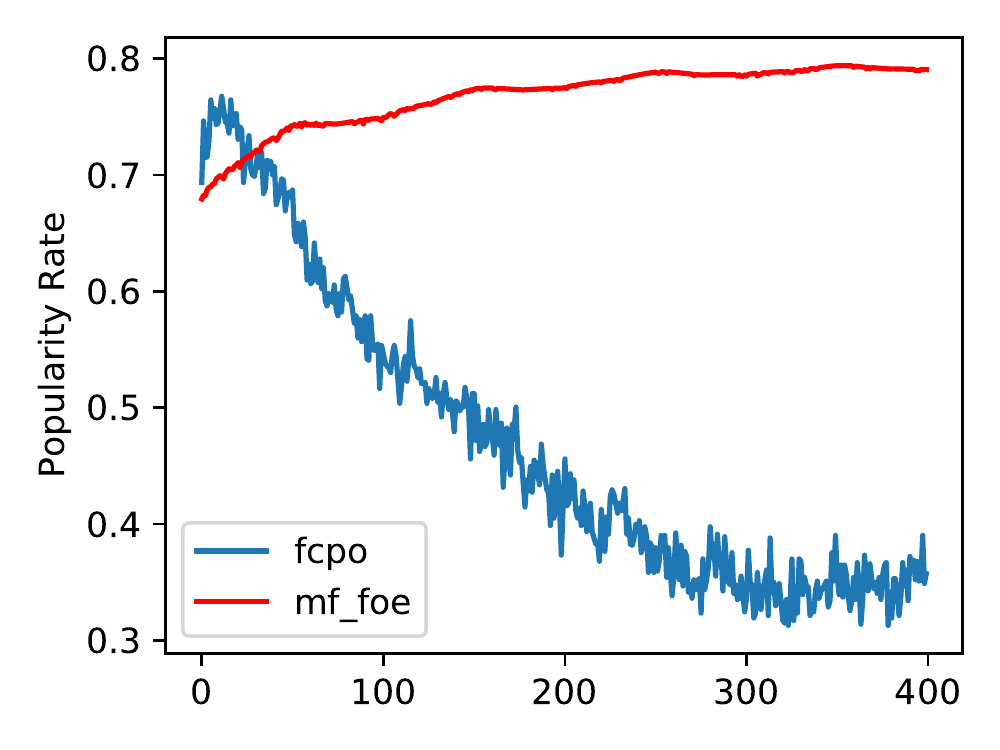}}
    \hspace{-10pt}
    \subfigure[Popularity Rate on Movielens1M]{\label{fig:ml1m_long_term_pr}
        \includegraphics[width=0.26\textwidth]{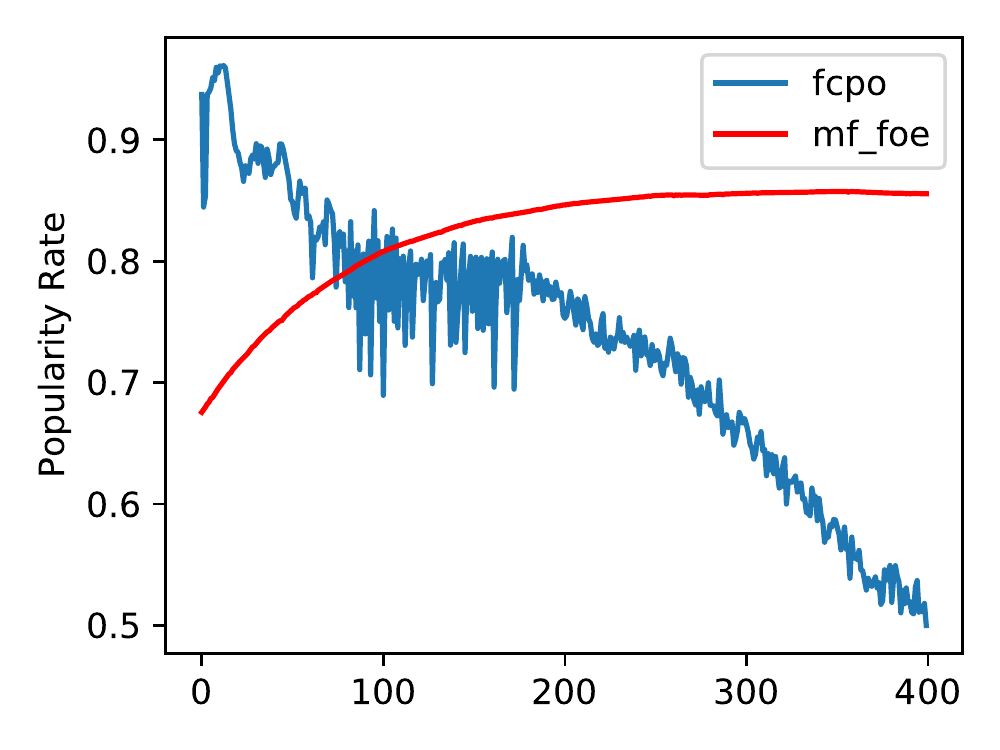}}

}
% \vspace{-10pt}
\caption{Long-term performance on Movielens100K (first column) and Movielens1M (second column).
X-axis is recommendation step, y-axis is the evaluated metric (first row: \textbf{NDCG}, second row: \textbf{Gini}, third row: \textbf{Popularity Rate}) on accumulated item exposure from beginning to current step.}
\label{fig:long_term}
% \vspace{-10pt}
\end{figure}

\subsection{Long-term Fairness in Recommendation} 

We compared FCPO with a static short-term fairness solution (i.e., MF-FOE) for 400 steps of recommendation.
For MF-FOE, we run 4 rounds of $K=100$ recommendations to let it capture the dynamics of the item group labels, while FCPO only needs to continuously run for 400 steps.
In other words, MF-FOE keeps the same item group labels for $K$ item recommendations and has to retrain its parameters after the labels updated at the end of each round.
As mentioned in section \ref{sec:experimental_setup}, FOE-based method becomes significantly time-consuming when dealing with large candidate item sets.
Thus, instead of doing whole item set fairness control, we first select the top $2K$ items as candidates, and then apply FOE to rerank the items and generate the final $K$ recommendations.

As shown in Fig. \ref{fig:ml100k_long_term_gini}, \ref{fig:ml1m_long_term_gini}, \ref{fig:ml100k_long_term_pr}, and \ref{fig:ml1m_long_term_pr}, when model convergences, MF-FOE performs much worse than FCPO on both Gini Index and Popularity Rate on two datasets.
% For these two fairness metrics, the 0-th entry before the first recommended item correspond to the feature of the original dataset, so the item exposure of MF-FOE exhibits a higher popularity rate and a minor improvement on GINI index from the original data.
Within each round of MF-FOE, fairness metrics quickly converges and they are further improved only when the item exposure information is updated.
On the contrary, since FCPO makes adjustment of its policy according to the fairness feedback, it can successfully and continuously suppress the fairness metric to a much lower value during testing.
As shown in Fig.\ref{fig:ml100k_long_term_pr} and Fig.\ref{fig:ml1m_long_term_pr}, due to this dynamic change of recommendation policy, FCPO exhibits greater fluctuation and unstable behavior than MF-FOE.
Though we kept skeptical whether the fairness performance gap between MF-FOE and FCPO will eventually vanish, we do observe that MF tends to much favor popular items than unpopular ones in Table \ref{tab:result}.
As a result, setting a very small $K$ (e.g. $K<20$) to speed up the recommendation could result in a candidate set filled with popular items and applying FOE becomes futile.
% Besides, the performance of MF-FOE m to further improve.
% By contrast, FCPO is aware of its fairness of each recommendation step and makes adjustment of its policy according to the fairness constraint.
% For each session, we update the item group with the item exposure from the previous session.
% The dynamics of item group are given by Figure \ref{fig:group_dynamics}.
Besides, the overall performance of MF-FOE -- especially on accuracy metrics (corresponding to Fig. \ref{fig:ml100k_long_term_ndcg} and \ref{fig:ml1m_long_term_ndcg}) -- is consistently outperformed by FCPO, which indicates that MF-FOE sacrifices the recommendation performance more than FCPO in order to control fairness.

\section{Conclusion and Future Work}
In this work, we propose to model the long-term fairness in recommendation with respect to dynamically changing group labels. We accomplish the task by addressing the dynamic fairness problem through a fairness-constrained reinforcement learning framework. 
% Specially, 
Experiments on standard benchmark datasets verify that our framework achieves better performance in terms of recommendation accuracy, short-term fairness, and long-term fairness.
% trade-off between recommendation performance and fairness performance from a short-term perspective, but also from a long-term perspective.
In the future, we will generalize the framework to optimize individual fairness constraints and other recommendation scenarios such as e-commerce recommendation and point-of-interest recommendation.
% generalize it to more fairness constraints.

% \newpage
% \balance
\bibliographystyle{ACM-Reference-Format}
\bibliography{paper}

\end{document}